\documentclass{JHEP3}
\usepackage[english]{babel}
\usepackage{amssymb}
\usepackage{amsfonts}
\usepackage{amsmath}
\usepackage{graphicx}
\usepackage{epsfig}
\usepackage{cite}
\newcommand{\cC}{{\cal C}}  
  
\newcommand{\cG}{{\cal G}}  
  
  \newcommand{\cL}{{\cal L}}
  \newcommand{\cN}{{\cal N}}
\newcommand{\cO}{{\cal O}}  \newcommand{\cP}{{\cal P}}

\newcommand{\bR}{{\mathbf R}}

\newcommand{\be}{\begin{equation}} \newcommand{\ee}{\end{equation}}
\newcommand{\bea}{\begin{eqnarray}} \newcommand{\eea}{\end{eqnarray}}
\newcommand{\beann}{\begin{eqnarray*}}  \newcommand{\eeann}{\end{eqnarray*}}
\newcommand{\bfig}{\begin{figure}} \newcommand{\efig}{\end{figure}}
\newcommand{\ba}{\begin{array}} \newcommand{\ea}{\end{array}}
\newcommand{\bcen}{\begin{center}} \newcommand{\ecen}{\end{center}}
\newcommand{\btab}{\begin{tabular}} \newcommand{\etab}{\end{tabular}}

\def\tr{\operatorname{tr\:}}

\newcommand{\vev}[1]{\left\langle{#1}\right\rangle}

\newtheorem{Proposition}{Proposition}[section]

\newtheorem{Theorem}{Theorem}[section]
\newtheorem{Lemma}{Lemma}[section]
\newtheorem{Corrolary}{Corrolary}[section]

\newcommand{\bp}{\begin{Proposition}}   \newcommand{\ep}{\end{Proposition}}
\newcommand{\bt}{\begin{Theorem}}   \newcommand{\et}{\end{Theorem}}
\newcommand{\bl}{\begin{Lemma}}     \newcommand{\el}{\end{Lemma}}
\newcommand{\bc}{\begin{Corrolary}} \newcommand{\ec}{\end{Corrolary}}
\title{Drag and jet quenching of heavy quarks in a strongly coupled {\boldmath $\cN=2^*$} plasma}
\author{Carlos Hoyos \\
  Department of Physics, \\
   University of Washington\\
  ~\,Seattle, WA 98915-1560, USA\\
  ~\,E-mail: \email{choyos@phys.washington.edu}}
 \abstract{
The drag of a heavy quark and the jet quenching parameter are studied in the strongly coupled ${\cal N}=2^*$ plasma using the AdS/CFT correspondence. Both increase in units of the spatial string tension as the theory departs from conformal invariance. The description of heavy quark dynamics using a Langevin equation is also considered. It is found that the difference between the  velocity dependent factors of the transverse and longitudinal momentum broadening of the quark admits an interpretation in terms of relativistic effects, so the distribution is spherical in the quark rest frame. When conformal invariance is broken there is a broadening of the longitudinal momentum distribution. This effect may be useful in understanding the jet distribution observed in experiments.
}


\begin{document}

\section{Motivation}

Experiments of heavy ion collisions at RHIC show a quantitative deviation from perturbative QCD predictions of heavy quark jet production \cite{rhic}. There are several indications that a deconfined state of matter that behaves as a strongly coupled plasma is formed at the center of the collisions.
For instance, hydrodynamic simulations using a very low viscosity have been quite successful in describing the observed elliptic flow at low transverse momentum \cite{hydroQCD}. Another example is the disappearance of back-to-back jets, that could be understood as the effect of a highly dissipative medium. Further evidence comes from lattice theory \cite{latticeEOS}, where it was observed that at the energy densities reached at RHIC the equation of state of QCD is still far from the free gas value.

Some of these observations agree with predictions of AdS/CFT duality \cite{firstadscft} that provides a holographic description of some strongly coupled gauge theories. For instance, the pressure of the strongly coupled $\cN=4$ super Yang-Mills theory was predicted to be about 75\% of the free gas \cite{AdSEOS}, which is similar to the results found in the lattice theory in the range of energies of RHIC. Another prediction was that the plasma shows a hydrodynamic behavior with a very small viscosity \cite{hydroAdS}. In the context of heavy quarks, a classical string computation in AdS/CFT gives predictions for the drag force \cite{Herzog:2006gh,CasalderreySolana:2006rq,Gubser:2006bz}, and there are different estimates for the energy loss of a fast particle moving through the medium. The various approaches involve studying string configurations corresponding to massless quarks \cite{Chesler:2008wd,Chesler:2008uy}, the jet quenching parameter from a dipole source \cite{Liu:2006ug} or the correlation functions of charged currents in the plasma \cite{Hatta:2008tx}. The behavior of the drag and jet quenching in a non-conformal theory will be the main interest of this paper.

Both the drag force and the jet quenching have been studied in a large number of situations \cite{Caceres:2006dj,Caceres:2006as,Matsuo:2006ws,Nakano:2006js,Talavera:2006tj,Gubser:2006qh,Bertoldi:2007sf,Cotrone:2007qa,Fadafan:2008gb,Fadafan:2008uv,Sadeghi:2008ws,Sadeghi:2008ci,Sadeghi:2009mp,Mia:2009wj,Giecold:2009wi,Horowitz:2009pw,Sadeghi:2009hh,Buchel:2006bv,Lin:2006au,Avramis:2006ip,Armesto:2006zv,Gao:2006uf,Argyres:2006yz,Gao:2007zx,Liu:2008tz,DeWolfe:2009vs,Gursoy:2009kk}, including some non-conformal theories. As opposed to the shear viscosity, these quantities do not show an universal behavior. Although according to the lattice results the equation of state of QCD in the RHIC range is not very far from a conformal theory, it is necessary to quantify how the deviation from conformal invariance will affect the energy loss of the jets. For this, one needs to use a framework that can connect smoothly with the conformal case. There have been several recent attempts in the context of AdS/QCD that show an interesting behavior \cite{Liu:2008tz,DeWolfe:2009vs,Gursoy:2009kk} , but it is certainly desirable to study a case where the deformation is fully understood. For this purpose, the $\cN=2^*$ theory \cite{Freedman:1999gp,Pilch:2000ue} is an ideal laboratory. Conformal invariance is explicitly broken by the introduction of a mass deformation in the $\cN=4$ theory. The ultraviolet fixed point is the $\cN=4$ theory, so the holographic correspondence is very well understood. The thermodynamic analysis \cite{Buchel:2003ah,Buchel:2007vy} shows that even when the equation of state starts deviating significantly from the conformal case there are no phase transitions. Therefore, one can tune the mass from zero to quite a large value and study the variation of the observables of interest.

Although the numerical results are given for the $\cN=2^*$ theory, the analytic results presented here do not depend on its particular properties, and should be general for a large class of models that describe scalar relevant deformations of a strongly coupled conformal theory in four dimensions. One should add that the deformation must have a dual description in terms of a scalar supergravity field, which it is not necessarily true for all of the possible relevant deformations in the field theory.

The paper is organized as follows. In section \ref{n2sec} there is a small review of the main features of the $\cN=2^*$ holographic dual. In section \ref{bhsec} some general properties of the black hole solutions and the thermodynamics of the $\cN=2^*$ theory are discussed. The choice of string configurations used to compute the different quantities is explained in section \ref{sec:averagewilson}. Section \ref{sigmassec} contains a derivation of the spatial string tension that will be used as reference scale. The next sections contain the computation of the drag coefficient and momentum broadening (section~\ref{dragsec}), and the jet quenching parameter (section~\ref{jetsec}). The numerical method used to calculate the values in the $\cN=2^*$ case is explained in the appendix \ref{app}. Results are summarized and discussed in section \ref{concsec}.

\section{Preliminaries}

\subsection{The supergravity dual of $\cN=2^*$ }\label{n2sec}

The contents of this section can be found in the references \cite{Pilch:2000ue,Buchel:2000cn,Evans:2000ct}, but are included here for completeness. The field content of $\cN=4$ can be grouped in $\cN=1$ superfields as a vector multiplet and three chiral superfields with scalar components
\begin{equation}
\Phi_i= X_{2 i -1}+i X_{2i},\ \ i=1,2,3\,.
\end{equation}
In the $\cN=2^*$ theory, two of the chiral fields $\Phi_1$ and $\Phi_2$ are grouped in a $\cN=2$ hypermultiplet and a mass is introduced in the Lagrangian. The remaining fields form an $\cN=2$ vector multiplet. The mass is introduced through the following relevant operators:
\begin{equation}
\begin{array}{rcl}
\cO_2 & =&\tr(-X_1^2-X_2^2-X_3^2-X_4^2+2 X_5^2+2 X_6^2), \\
\cO_3 &= &\tr(\lambda_1\lambda_1+\lambda_2\lambda_2)+({\rm scalar \; trilinear})+h.c.\,.
\end{array}
\end{equation}
where $\lambda_i$ are the fermionic components of the multiplets. This choice of operators correspond to the $SO(6)$ representations ${\bf 20'}$ and ${\bf 10}+{\bf \overline{10}}$, that map to two different scalar fields in the holographic dual, $\alpha$ and $\chi$. The last can couple to the scalar singlet mass operator,
\begin{equation}
\cO_1=\tr(X_1^2+X_2^2+X_3^2+X_4^2+X_5^2+X_6^2)\,,
\end{equation}
that is not BPS protected and has no associated supergravity field. The mass term comes from the combination
\begin{equation}
\cO_1-{1\over 2}\cO_2={3\over 2}\tr (X_1^2+X_2^3+X_3^3+X_4^2 )\,.
\end{equation}

The dual theory is described by solutions of five-dimensional $\cN=8$ supergravity with the scalar fields turned on. This is a consistent truncation of type IIB supergravity in ten dimensions, and it is possible to uplift the five-dimensional solutions to ten dimensions, although this is far from trivial. In the zero temperature case the five-dimensional metric can be put in the form
\begin{equation}
ds_{4,1}^2=e^{2A(r)}(- dt^2+d\vec{x}^2)+d r^2\,.
\end{equation}
Thanks to supersymmetry, it is possible to obtain the function $A(r)$ and the scalar fields $\alpha(r)$, $\chi(r)$ as solutions of a system of first order differential equations.
\begin{equation}\label{susyeqs}
{dA\over dr} = -{g\over 3} W, \ \ {d\alpha\over d r} = {g\over 4} {\partial W\over \partial \alpha}, \ \ {d\chi\over d r} = {g\over 4} {\partial W\over \partial \chi}\,.
\end{equation}
where
\begin{equation}\label{superpotential}
W=-e^{-2\alpha/\sqrt{3}}-{e^{4\alpha/\sqrt{3}}\over 2} \cosh(2\chi)\,.
\end{equation}
is the superpotential and $g$ is the five-dimensional gauge coupling. The value of $g$ sets the radius of curvature $L=2/g$, related to the 't Hooft coupling in the dual theory via $L^4=g_{YM}^2 N (\alpha')^2 = \lambda (\alpha')^2$\,.

The zero temperature theory has an interesting structure that in the gauge theory can be formulated in terms of the Coulomb branch of the moduli space. On the gravity side the supergravity solution has a singularity at the origin of the space. This singularity is of the good kind \cite{Gubser:2000nd} and from an analysis of D-brane and string probes in the geometry, it is possible to relate it to a particular configuration in the Coulomb branch of the $\cN=2^*$ theory where all dyonic states become massless on a ring (the enhan\c con) around the origin of moduli space \cite{Buchel:2000cn,Evans:2000ct}. At finite temperature the Coulomb branch is lifted and there is no enhan\c con singularity in the geometry.

At non-zero temperature supersymmetry is broken and it is necessary to use the second order equations of motion \cite{Buchel:2003ah,Buchel:2007vy}.
\begin{equation}\label{scalareqs}
\square \alpha = {1\over 2}{\partial \cP \over \partial \alpha}, \ \ \ \square \chi = {1\over 2}{\partial \cP \over \partial \chi}\,.
\end{equation}
The potential for the scalar fields is
\begin{equation}
\cP = {g^2\over 16} \left[\left({\partial W\over \partial \alpha} \right)^2+\left({\partial W\over \partial \chi} \right)^2 \right]-{g^2\over 3} W^2\,.
\end{equation}
The metric can be written as
\begin{equation}\label{5dmetric}
ds_{4,1}^2=e^{2A(r)}\left( -e^{2 B(r)} dt^2+d{\vec x}^2\right)+d r^2\,,
\end{equation}
and the functions $A(r)$ and $B(r)$ can be obtained from the Einstein equations
\begin{equation}\label{einsteineqs}
{1\over 4}R_{\mu\nu}= T_{\mu\nu}-{1\over 3} g_{\mu\nu} T^\rho_\rho
\end{equation}
where
\begin{equation}
T_{\mu\nu}=\partial_\mu\alpha \partial_\nu \alpha+\partial_\mu\chi \partial_\nu \chi-{1\over 2} g_{\mu\nu}\left[\partial_\rho\alpha \partial^\rho \alpha+\partial_\rho\chi \partial^\rho \chi+ \cP\right]\,.
\end{equation}
is the energy-momentum tensor of the scalar fields.

From the equations of motion one can deduce the asymptotic behavior of the scalar fields. As $r\to\infty$
\begin{equation}\label{scalarboundexp}
\chi  =  k e^{-r/L}(1+\dots), \ \ \rho \equiv e^{\alpha/\sqrt{3}} =  1-{2\over 3}k^2 {r\over L} e^{-2 r\over L}+\dots\,,
\end{equation}
where $k=mL$ is the mass of the hypermultiplets. This radial dependence corresponds to scalar fields with masses
\begin{equation}
m_\chi^2L^2= -3, \ \ m_\alpha^2L^2 =-4.
\end{equation}
This matches the relation between the mass of a scalar field and the conformal dimension of the dual operator, $m_s^2L^2=\Delta(\Delta-4)$, with  $\Delta=3$ for the $\cO_3$ operator and $\Delta=2$ for the $\cO_2$ operator. At the horizon $r\to 0$ the scalar fields have a constant value
\begin{equation}\label{scalarexphoriz}
\chi = \chi_0+\chi_1 r^2+\dots , \ \ \rho = \rho_0+\rho_1 r^2+\dots\,.
\end{equation}

As was mentioned before, it is possible to uplift the five-dimensional solutions to a full ten-dimensional geometry. In the Einstein frame it looks
\begin{equation}\label{10dmetric}
ds_{10}^2=\widetilde\Omega^2 ds_{4,1}^2+ds_5^2\,,
\end{equation}
where
\begin{equation}
ds_5^2={L^2 \widetilde\Omega^2 \over \rho^2}\left( c^{-1} d\theta^2 +\rho^6 \cos^2\theta\left({\sigma_1^2\over c X_2}+{\sigma_2^2+\sigma_3^2\over X_1}\right)+\sin^2\theta {d\phi^2\over X_2}\right)
\end{equation}
with the warp factor $\widetilde \Omega$, $c$  and $\rho$ defined as
\begin{equation}\label{warp}
\widetilde\Omega^2={(c X_1 X_2)^{1/4}\over \rho}, \ \ c=\cosh(2\chi), \ \ \rho=e^{\alpha/\sqrt{3}}\,.
\end{equation}
The scalar functions are
\begin{equation}
\begin{array}{rcl}
X_1 & = & \cos^2\theta+\rho^6 c \sin^2\theta\,, \\
X_2 & = & c \cos^2\theta+\rho^6 \sin^2\theta\,.
\end{array}
\end{equation}
The one-forms $\sigma_i$ are the $SU(2)$ left-invariant forms satisfying $d\sigma_i = \epsilon_{ijk} \sigma_j\wedge \sigma_k$.

For the computation of string solutions, one also has to take into account the dilaton
\begin{equation}\label{dilaton}
e^{-\varphi}={1\over 2}\left( \left({c X_1\over X_2}\right)^{1/2}+\left({c X_1\over X_2}\right)^{-1/2}\right)\,.
\end{equation}
The axion and the three and five form fields are also present in the geometry, but they will not be important for the analysis presented here. This is the case because the components of the NS flux along the directions of the string probes studied here vanish. In other cases the situation can be different, an example of this would be the non-relativistic backgrounds of \cite{Adams:2008wt}. This would affect for instance the considerations of section \ref{sec:averagewilson}.

\subsection{Black hole properties and thermodynamics}\label{bhsec}

Some general properties will be presented here. For concrete calculations the equations are solved numerically. Details are explained in the appendix \ref{app}.

First recall the form of the five-dimensional black hole metric
\begin{equation}
d s_{4,1}^2 = e^{2 A(r)} (-e^{2 B(r)} dt^2+d x^2)+dr^2\,.
\end{equation}
For asymptotically $AdS$ spaces, the behavior as $r\to \infty$ is
\begin{equation}
A(r)\to {r\over L}+A_0\,,
\end{equation}
and
\begin{equation}
B(r)\to 0\,.
\end{equation}
The constant $A_0$ is set to zero by a rescaling of the spacetime coordinates. It is always possible to choose a coordinate system where locally the metric component $g_{rr}=1$, so this form of the metric is completely general for a geometry that has spatial rotational invariance.

The horizon is at $r=0$, where the blackening function $e^{2B}$ vanishes. For a regular horizon the metric functions have the following expansions
\begin{equation}\label{expanb0}
e^{B(r)} = b_0 {r\over L} \left(1+b_1 {r^2\over L^2}+\dots\right)\,,
\end{equation}
\begin{equation}\label{expan0}
e^{A(r)} = a_0 \left(1+ a_1 {r^2\over L^2}+\dots\right)\,.
\end{equation}
The smoothness of the Euclidean solution and the area of the horizon determine the temperature and entropy density
\begin{equation}
T={a_0 b_0\over 2\pi L},  \ \ S= {a_0^3\over 4 G_5}\,.
\end{equation}
($G_5$ is the 5d Newton constant). Although in the uplift to the ten dimensional metric \eqref{10dmetric} the internal space and the conformal factor depend on the scalars, one can show that they do not affect the values of $T$ or $S$\footnote{The area of the horizon just gets a factor that is the volume of the undeformed $S^5$.}.

In $AdS_5$ (${\cal N}=4$ theory), the values are
\begin{equation}
a_0={u_H\over L}, \ \  b_0=2\,,
\end{equation}
where $u_H$ is the horizon radius in Schwarzschild coordinates and $b_0^2=4$ is related to exponent in the blackening function  $f(u)=1-u_H^4/u^4$, that in the $AdS$ space is fixed by the number of dimensions.

In the general case, the entropy can be cast in a more useful form in terms of the $\cN=4$ entropy. Using
\begin{equation}
{L^3\over 4 G_5} = {N^2\over 2\pi}\,,
\end{equation}
the entropy density is
\begin{equation}\label{entropy}
S={4\pi^2\over b_0^3} N^2 T^3 = {8\over b_0^3} S_{{\cal N}=4}\,.
\end{equation}
So the ratio $2/b_0$ controls the deviation from the conformal case.

One of the Einstein equations can be integrated, giving a relation that will be useful for the computation of the drag coefficient. Notice that for scalar fields that depend only on the radial coordinate, the combination
\begin{equation}
g_{ii} R_{00}-g_{00} R_{ii}=0, \ \ \  i=1,2\; {\rm or}\; 3\,,
\end{equation}
is independent of the scalar fields. This gives the equation
\begin{equation}\label{constEinst}
B''+(4A'+B')B'=0\ \ \Rightarrow \ \ \ln B'+ 4A+B=\,{\rm const}\,.
\end{equation}
This formula is true for any Lorentz-invariant deformation involving scalar fields, not only for the $\cN=2^*$ theory.

These formulas are quite general, but in order to describe the thermodynamic behavior of the $\cN=2^*$ plasma one has to solve the equations of motion and find the right geometries. This was done in \cite{Buchel:2007vy,Buchel:2003ah} and later on, hydrodynamic properties of the $\cN=2^*$ theory were also studied in \cite{Buchel:2004hw,Benincasa:2005iv,Buchel:2007mf,Buchel:2008uu}. The free energy was computed numerically up to values $m/T\sim 7$. A good approximation is given by the fit
\begin{equation}
F_{\cN=2^*}\simeq F_{\cN=4} e^{-{m/(7 T)}}
\end{equation}
For a value $m/T\simeq 6$, the speed of sound was found to be $v_s^2 \simeq 0.825 \,v_{CFT}^2$, with $v_{CFT}^2 =1/3$. This implies that the equation of state of the $\cN=2^*$ plasma is quite close to the conformal case up to values of the mass $m/T \sim 7$. If the bulk viscosity over shear viscosity ratio $\zeta/\eta$ approximately saturates the bound observed in \cite{Buchel:2007mf}
\begin{equation}
{\zeta \over \eta} \geq 2\left({1\over 3} - v_s^2\right) \,,
\end{equation}
then, at $m/T\simeq 6$, $\zeta/\eta\simeq 0.12$ is quite small, but starts to be comparable to the shear viscosity. In the numerical evaluation of quantities, the mass will be taken up to larger values $m/T\sim 17$, so the theory is in the region where deviations from conformality should be substantial.

\subsection{Averaged Wilson loops}\label{sec:averagewilson}

The analysis of the energy loss of heavy quarks in the plasma involves the computation of some Wilson loop configurations. In $\cN=4$ it is possible to use a classical string configuration in the holographic dual to evaluate the expectation value of a generalized version of a Wilson loop, involving the gauge and the scalar fields
\cite{Maldacena:1998im}
\begin{equation}\label{wilson}
\vev{W(\cC,\theta)} =\vev{e^{i\int_\cC d \tau\,  (A_\mu \dot x^\mu+\theta^I X_I \sqrt{\dot x^2})}}\,.
\end{equation}
where $x^\mu(\tau)$ parameterizes the contour $\cC$ of the Wilson loop and the $\theta^I$ $I=1,\dots,6$ are components of a unit vector in $\bR^6$, so they parameterize a five-sphere $S^5$. In the holographic dual the $S^5$ is realized geometrically, and the generalized Wilson loops can be described using classical string configurations that sit at a point of the $S^5$. In the $\cN=4$ theory the $SO(6)$ R-symmetry that acts on the $S^5$ is unbroken, so the values of the Wilson loops are independent of the value of $\theta^I$.

In the $\cN=2^*$ theory the moduli space is partially lifted by the masses, so the $S^5$ in the holographic dual is deformed. The $SO(6)\simeq SU(4)$ symmetry is broken to a $SU(2)\times U(1)$ subgroup. In this case one should be careful with the choice of Wilson loop, since the expectation values now depend on the values of $\theta^I$. Other situations where this can happen include the R-charged black holes, as shown in \cite{Herzog:2007kh}. A possible definition of a Wilson loop that is independent of the $\theta^I$'s is an averaged Wilson loop
\begin{equation}
\vev{\overline W(\cC)} = {1\over V(S^5)} \int_{S^5} d \Omega_5(\theta) \vev{W(\cC,\theta)}\,.
\end{equation}
Where the average is over the parameters $\theta^I$ that define the Wilson loop in the field theory. This corresponds to an object with zero R-charge, so it is closer to the properties of a Wilson loop involving only gauge fields. A similar kind of average was considered in \cite{Dorn:2007zy} for Wilson lines in the $\cN=4$ theory with dependence on the internal directions, as a way to estimate the quark-antiquark potential for QCD.

Clearly in the $\cN=4$ theory the averaged Wilson loop has the same expectation value as \eqref{wilson}. Although it is possible to find an explicit expression for $\vev{\overline W(\cC)}$, one can use a saddle point approximation instead. For this, one writes the Wilson loop as
\begin{equation}
\vev{W(\cC,\theta)}= e^{i S(\cC,\theta)} \ \ {\rm or} \ \ \vev{W(\cC,\theta)}= e^{-S(\cC,\theta)}\,,
\end{equation}
depending on whether the Wilson loop is timelike or spacelike. The action $S(\cC,\theta)$ computed in holographic models has a large $\sqrt{\lambda}$ factor, so in the strong coupling limit the integral over the $S^5$ volume will be dominated by minimal action configurations\footnote{Strictly speaking, for timelike Wilson loops the condition is that the action is extremized, but in the cases studied here the physically sensible saddle point corresponds to the position on the $S^5$ where the action is minimized.}
\begin{equation}
\vev{\overline W(\cC)} \simeq \vev{W(\cC,\theta_{\rm min})} \,.
\end{equation}
where $\theta_{\rm min}$ is the value that minimizes the action. There is actually a full set of values related by the remaining $SU(2)\times U(1)$ symmetry, but as for the $\cN=4$ case, it is enough to take one representative.  Notice that this formula would be valid only as long as the string configuration is at a fixed point in the compact space or moves along an isometric direction. That would not be the case if for instance an NS flux induces a force on the string, as was mentioned in section \ref{n2sec}. The minimal action configurations at a fixed point will be the ones used to compute the string tension, drag and jet quenching.

A generalized Wilson loop \eqref{wilson} has a holographic description in the semiclassical limit as a classical string configuration in ten dimensions. The string is a two-dimensional surface with a boundary at $r\to\infty´$ that is the contour of the Wilson loop $\cC$. The configuration is determined by the equations derived from the Nambu-Goto action in the Einstein frame,
\begin{equation}
S_{NG}={1\over 2\pi\alpha'} \int d^2 \sigma \cL= {1\over 2\pi\alpha'} \int d^2 \sigma\; \Omega^2 \sqrt{g_2}, \ \ \ \Omega^2\equiv e^{\varphi/2}\widetilde\Omega^2\,,
\end{equation}
where $g_2$ is the pullback of the five-dimensional metric \eqref{5dmetric}, $\varphi$ is the dilaton \eqref{dilaton} and $\tilde \Omega$ is the warp factor \eqref{warp}. From \eqref{scalarboundexp}, the factor $\Omega^2 \to 1$ as $r\to \infty$. On the other hand, when $r\to 0$ the scalar fields tend to a constant \eqref{scalarexphoriz} and the factor $\Omega^2$ in the string action depends on the position of the string in the internal space.

String configurations with minimal action are localized at $\theta=\pm \pi/2$, that corresponds to the part of the geometry associated to the $\cN=2$ moduli space. In this case the factor $\Omega^2$ is minimized for all values of $r$,
\begin{equation}
\Omega^2 =\rho^2  \sqrt{2\cosh^2 (2\chi)\over 1+\cosh^2(2\chi)}\,.
\end{equation}
So there will be a dependence on the value of the scalar fields. In the $r\to 0$ limit, the factor becomes a constant:
\begin{equation}
\Omega_0^2= \rho_0^2  \sqrt{2\cosh^2(2\chi_0)\over 1+\cosh^2(2\chi_0)}\;.
\end{equation}

\section{Spatial string tension}\label{sigmassec}

Spatial Wilson loops in the deconfined phase show an area law behavior when the size of the Euclidean time is very small compared to the size of the loop. It is interesting to compute the spatial string tension using holography for two different reasons. In the first place, it is a physical quantity that one may compare directly with results from lattice computations \cite{latticeEOS,Kiskis:2009xj} (see \cite{Schroder:2005zd} for a comparison with different methods). In a phenomenological application to the QCD quark-gluon plasma, one could fix the string tensions of lattice QCD and of the holographic model to be equal and extract the values of other parameters, in the spirit of previous works, as for example in \cite{Gubser:2006qh,CasalderreySolana:2006rq}. The second reason, and the one that motivates its calculation here, is that it has the same parametric dependence on the 't Hooft coupling as the drag force coefficient and the jet quenching, so it is canceled out in the ratio. In principle one could accomplish the same by taking the ratio with the values of the conformal theory, but notice that this does not give useful information about intrinsic properties of the theory. In the next sections the values of the drag coefficient and jet quenching parameter will be expressed in terms of the string tension.

A spatial Wilson loop can be described holographically by a classical string configuration ending on the appropriate contour at the boundary. The contour of the Wilson loop is chosen to be rectangular in the $x$ and $y$ directions. The length in the $y$ direction $L_y$ is taken to be very large, so it is a good approximation to neglect any dependence on the $y$ coordinate. The choice of worldsheet coordinates is $y=\tau$ and $r=\sigma$, and the profile of the string is determined by a function $x(r)$. The string worldsheet Lagrangian is
\begin{equation}
\cL=\Omega^2 e^A\sqrt{1+e^{2A}{x'}^2}\,.
\end{equation}
From this, the following equation of motion is obtained
\begin{equation}
x'={c e^{-A}\over \sqrt{\Omega^4e^{4A}-c^2}}\,.
\end{equation}
Choosing the integration constant $c=\Omega^2(r_0) e^{2 A(r_0)}$ implies that the string profile ends at $r_0$. The length in the $x$ direction and the regularized action of the Wilson loop\footnote{Obtained by subtracting the straight configuration $x=0$.} are
$$
{L_x\over 2} =\int_{r_0}^\infty dr\; {c e^{-A}\over \sqrt{\Omega^4e^{4A}-c^2}}\,,
$$
\begin{equation}
 {S\over 2}={L_y\over 2 \pi \alpha'} \left(\int_{r_0}^\infty dr\;\Omega^2 e^{ A}\left[{\Omega^2 e^{2 A}\over \sqrt{\Omega^4 e^{4A}-c^2}}-1 \right]-\int_0^{r_0} dr\; \Omega^2e^A\right)\,.
\end{equation}
The length grows as $r_0\to 0$. In that regime, the contributions to the integrals from $r\to\infty$ are small, since the integrands decay exponentially and the integrand has a square root divergence at $r_0$. One can use the approximation $e^{A(r)}\simeq e^{A(r_0)}$. Defining $\tilde A(r)=A(r)-A(r_0)$ and $\tilde \Omega={\Omega(r)\over \Omega(r_0)}$,
$$
{L_x\over 2} \simeq {1\over a_0} \int_{r_0\to 0}^\infty dr \;{ 1\over \sqrt{{\tilde \Omega}^4 e^{4\tilde A(r)}-1}}\,,
$$
\begin{equation}
{S\over 2}\simeq {\Omega_0^2a_0 L_y\over 2\pi \alpha'} \int_{r_0\to 0}^\infty dr\;{1\over \sqrt{{\tilde \Omega}^4e^{4\tilde A(r)}-1}}\,.
\end{equation}
Remember that $a_0=e^{A(0)}$, as defined in \eqref{expan0}. At large distances $L_x\gg 1/T$ the Wilson loop follows an area law
\begin{equation}
S \sim \Omega_0^2 {a_0^2 \over 2\pi \alpha'} L_x L_y = \left({2\over b_0}\right)^2 \Omega_0^2 {\pi \sqrt{\lambda}T^2\over 2} L_x L_y \equiv \sigma_s L_x L_y\,.
\end{equation}
In terms of the $\cN=4$ theory result, the string tension is
\begin{equation}\label{stringt}
{\sigma_s\over \sigma_{\cN=4} }= \left({2\over b_0}\right)^2 \Omega_0^2\,.
\end{equation}
In figure \ref{stringtension} this approximate formula in the $\cN=2^*$ theory is compared with a direct evaluation using the string solutions for several values of $m/T$. The agreement is good only for small masses, although it captures the right tendency of the string tension to decrease as the mass increases.

\FIGURE[ht!]{
\includegraphics[width=12cm]{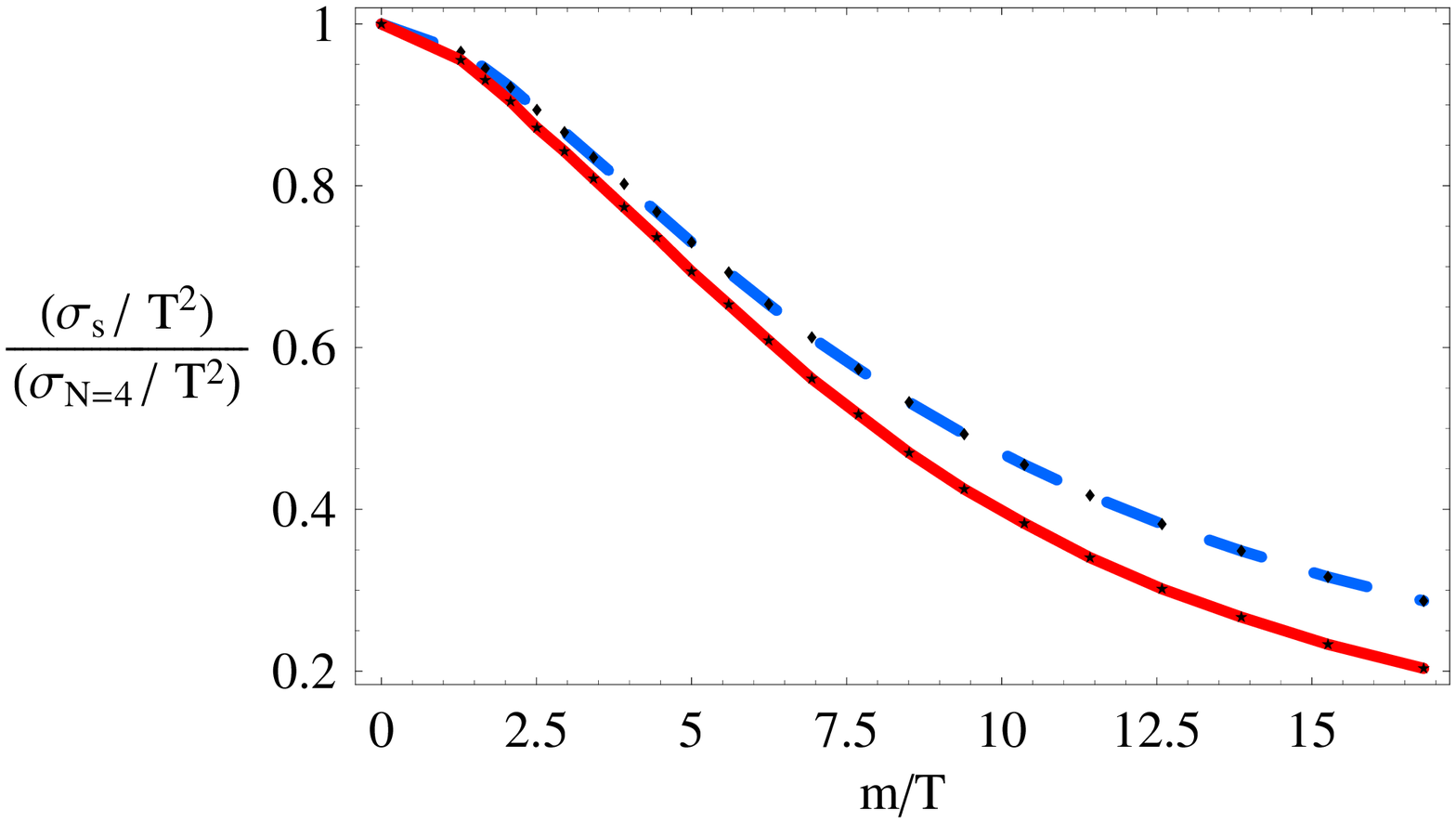}
\caption{\label{stringtension} Spatial string tension of $\cN=2^*$ theory for different values of the mass deformation. The dashed line corresponds to the approximate formula \eqref{stringt} and the solid line to a numerical evaluation using the string solution.}
}

\section{Drag force and Langevin equation}\label{dragsec}

In the formation of the quark-gluon plasma, soft light quarks and gluons thermalize very fast but heavy quarks remain out of equilibrium for a relatively much longer time if their mass is much larger than the temperature $M/T\gg 1$. At weak coupling the quarks lose energy through bremsstrahlung and collision processes, that can be described within the framework of kinetic theory. The momentum change of the heavy quark may be described by a Langevin equation. A quark with spatial momentum $p_i$ moving through the plasma experiences a force
\cite{Moore:2004tg}
\begin{equation}\label{langevin}
{d p_i\over dt}=-\eta_D p_i +\zeta^L_i(t)+\zeta^T_i(t)\,.
\end{equation}
This is a phenomenological description where the effect of the medium has been separated in two parts, the drag $\sim \eta_D$ and a stochastic contribution $\zeta^T$, $\zeta^L$. The drag force opposes to the movement of the quark $-\eta_D p_i$, where $\eta_D$ is the drag coefficient. The term coming from random collisions with the components of the plasma $\zeta^L_i$ and $\zeta^T_i$ broadens the momentum distribution in the longitudinal and transverse directions. Usually the random components are taken as Gaussian white noise, but in general the momentum transfer coefficients  $\kappa_L$ and $\kappa_T$ can depend on the momentum
\begin{equation}\label{zeta2point}
\vev{\zeta_i^L(t)\zeta_j^L(t')}= {p_i p_j\over p^2} \kappa_L(p) \delta(t-t'), \ \ \ \vev{\zeta_i^T(t)\zeta_j^T(t')}= \left(\delta_{ij} -{p_i p_j\over p^2}\right) \kappa_T(p) \delta(t-t').
\end{equation}
In principle, the drag coefficient can also depend on the momentum $\eta_D=\eta_D(p)$. The fluctuation-dissipation theorem relates the zero-momentum values of the drag coefficient and the momentum transfer $\kappa_L(0)=\kappa_T(0)=\kappa$, giving the Einstein relation
\begin{equation}\label{einstrel}
\eta_D(0)={\kappa \over 2 M T}\,.
\end{equation}

The momentum transfer also determines the diffusion of the quark through the plasma. Consider a quark at rest at $t=0$ and $x=0$. The mean squared position of the quark at later times will be
\begin{equation}
\vev{x_i(t) x_j(t)}=2 D t \delta_{ij}, \ \ \ D={2 T^2 \over \kappa}\,,
\end{equation}
where $D$ is the diffusion coefficient.

At strong coupling there is no well-defined kinetic description of the thermal medium, so it is not clear that a description in terms of the Langevin equation should work. Remarkably, in the holographic description of a heavy quark in the $\cN=4$ plasma, a drag force of the form appearing in \eqref{langevin} with a {\em constant} drag coefficient was found \cite{Herzog:2006gh,CasalderreySolana:2006rq,Gubser:2006bz}. Later a computation of the momentum transfer showed that it depends on the momentum and that the Einstein relation \eqref{einstrel} holds in the limit of low velocities \cite{CasalderreySolana:2007qw,Gubser:2006nz}.

\subsection{Holographic computation of the drag}\label{holodrag}

A heavy quark moving through the plasma is described by a trailing string configuration in the black hole background \cite{Herzog:2006gh,CasalderreySolana:2006rq,Gubser:2006bz}. The profile moves at constant velocity $v$ in the $x$ direction and it is extended from the boundary to the horizon in the radial direction $r$ and from $x=v t$ to $x\to-\infty$ in the spatial direction. Identifying the time $t$ and $r$ with the worldsheet coordinates, the profile is described by a function
\begin{equation}\label{dragprof}
x(r,t)=v t+\xi(r)\,,
\end{equation}
and by the position in the internal space in the ten-dimensional geometry. As was argued in section \ref{sec:averagewilson}, a reasonable choice is to pick a point that maps into the moduli space of the $\cN=2$ theory. Introducing \eqref{dragprof} in the Nambu-Goto action, the Lagrangian for $\xi(r)$ is
\begin{equation}\label{draglag}
{\cal L}=-\sqrt{-g}=- \Omega^2 e^A\left(e^{2B}-v^2+e^{2 A+2B} \xi'^2\right)^{1/2}\,.
\end{equation}

The drag force in the $x$ direction is given by the momentum flow along the string. Labeling $p_x$ the momentum in the $x$ direction, $G_{\mu\nu}$ the components of the background metric  and $g_{ab}$  the components of the induced metric on the worldsheet, the momentum flow is
\begin{equation}\label{pflow}
{d p_x\over dt}={1\over 2 \pi \alpha'}\sqrt{-g} G_{xx} g^{rr} \partial_r x(r,t) = {1\over 2\pi\alpha' \sqrt{-g}} \Omega^4 e^{4A+2B} \xi'=-{\pi_\xi\over 2\pi\alpha'}\,.
\end{equation}
Where $\pi_\xi$ is the canonical momentum
\begin{equation}
\pi_\xi=-{\partial {\cal L} \over \partial \xi'} = {\Omega^2 e^{3 A+2 B} \xi'\over \left(e^{2B}-v^2+e^{2 A+2B} \xi'^2\right)^{1/2}} \,.
\end{equation}
From this expression one finds the following equation for the profile
\begin{equation}\label{eomxi}
\xi' =\pi_\xi e^{-A-B} \sqrt{e^{2 B}-v^2\over \Omega^4 e^{4A+2B}-\pi_\xi^2}\,.
\end{equation}
In order to have a profile that can go all the way to the horizon, it is necessary that numerator and denominator flip sign at the same point. This imposes a condition on the canonical momentum
\begin{equation}\label{pixi}
\pi_\xi^2=\left. \Omega^4 e^{4A+2B}\right|_{v^2=e^{2B}}=\left. v^2\Omega^4 e^{4 A}\right|_{v^2=e^{2B}}\,.
\end{equation}
Using \eqref{pixi} in \eqref{pflow}, one can deduce the value of the drag force. For $v>0$, $\pi_\xi>0$,
\begin{equation}\label{stringdrag}
{d p_x\over dt}=-{\pi_\xi\over 2\pi\alpha'} = -\left. {v \Omega^2 e^{2 A}\over 2\pi\alpha'}\right|_{v^2=e^{2B}}\,.
\end{equation}

The expression \eqref{stringdrag} leads to the known result in the ${\cal N}=4$ theory,
\begin{equation}
{d p_x\over dt}=-{\pi\sqrt{\lambda}\over 2} T^2 {v\over\sqrt{1-v^2}}=-{\pi\sqrt{\lambda}\over 2} T^2 {p_x\over M}\equiv -\mu_{{\cal N}=4} {p_x\over M}\,,
\end{equation}
where a factor of the mass is extracted from the drag coefficient $\mu\equiv \eta_D M$ for convenience. Notice that $\mu_{{\cal N}=4}$ is independent of the velocity. In the general case this is not true, and it is interesting to extract the velocity dependence.

In terms of the functions appearing in the metric, the drag coefficient becomes
\begin{equation}\label{dragcoef}
\mu =-{M\over p_x} {d p_x\over d t} = {1\over 2\pi \alpha'}\left. (1-e^{2B})^{1/2} \Omega^2 e^{2 A}\right|_{v^2=e^{2B}}\,.
\end{equation}
At large velocities $v\to 1$, the value of the drag coefficient depends on the asymptotic behavior of the solution $r\to \infty$. The scalar fields do not affect to the result, since $\Omega^2\to 1$. The leading terms in the metric are
\begin{equation}\label{rinfexp}
e^{2 B} =1-\tilde\mu_1^2 e^{-4 r/L}+\dots\,, \ \ \ e^A =e^{r/L}(1+\dots)\,.
\end{equation}
where the constant $\tilde\mu_1$ corresponds precisely to the value of the drag coefficient up to $\alpha'$ factors. One can extract the value of $\tilde\mu_1^2$ using the relation \eqref{constEinst}. Comparing the $r\to \infty$ \eqref{rinfexp} and $r\to 0$ \eqref{expan0} limits, this gives
\begin{equation}
\ln(2\tilde\mu_1^2) = \ln b_0+4 \ln a_0\,,
\end{equation}
so the drag coefficient for ultrarelativistic quarks is
\begin{equation}\label{mu1}
\mu_{v\to 1}^2={b_0 a_0^4\over 2(2\pi\alpha')^2}\,.
\end{equation}
Contrary to previous claims the drag coefficient is sensitive to the infrared physics even at large velocities. Technically, the high-velocity drag $\tilde\mu_1^2$ appears in the expansion of the metric \eqref{rinfexp} as a normalizable term so it has a similar status to an expectation value, it should be possible to relate it to transport coefficients of the plasma.

At small velocities $v\to 0$, the drag coefficient is determined by the expansions at $r\to 0$ \eqref{expan0}
\begin{equation}
\mu_{v\to 0}^2={\Omega_0^4 a_0^4\over (2 \pi\alpha')^2}\,.
\end{equation}
The ratio depends on $b_0$ and the value of the scalar fields at the horizon
\begin{equation}\label{relativedrag}
\left({\mu_{v\to 0}\over \mu_{v\to 1}} \right)^2 = {2\over b_0} \Omega_0^4\,.
\end{equation}
The drag coefficient can be written in terms of the ${\cal N}=4$ coefficient:
\begin{equation}
\mu_{v\to 1}^2= {(2\pi L T)^4\over 2 (2\pi\alpha')^2 b_0^3}=\left({2\over b_0} \right)^3 \mu_{{\cal N}=4}^2\,,
\end{equation}
hence
\begin{equation}\label{mu0}
\mu_{v\to 0}^2= \left({2\over b_0} \right)^4 \Omega_0^4 \; \mu_{{\cal N}=4}^2\,.
\end{equation}
Using \eqref{stringt}, the drag coefficient would be
\begin{equation}\label{dragstunits}
\mu_{v\to 0}= \sigma_s, \ \ \mu_{v\to 1} = \sqrt{b_0\over 2} {1\over \Omega_0^2}\,\sigma_s\,.
\end{equation}
For $\cN=4$, this implies that $\mu=\sigma_s$, as was pointed out in \cite{Sin:2006yz}. However, this relation does not seem to hold in the non-conformal cases, because of the deviation of the string tension from the analytic result.
The numerical value of the drag coefficient in string tension units as a function of the velocity for several mass deformations has been plotted in figure \ref{dragfig}. Formula \eqref{relativedrag} is also compared with the numerical results, with very good agreement. The main effect of breaking conformal invariance is that the drag coefficient depends on the velocity, in the $\cN=2^*$ theory it is larger at higher velocities. Similar behavior was observed in other non-conformal theories \cite{Caceres:2006as}.

\FIGURE[ht!]{
\begin{tabular}{cc}
\includegraphics[width=7.25cm]{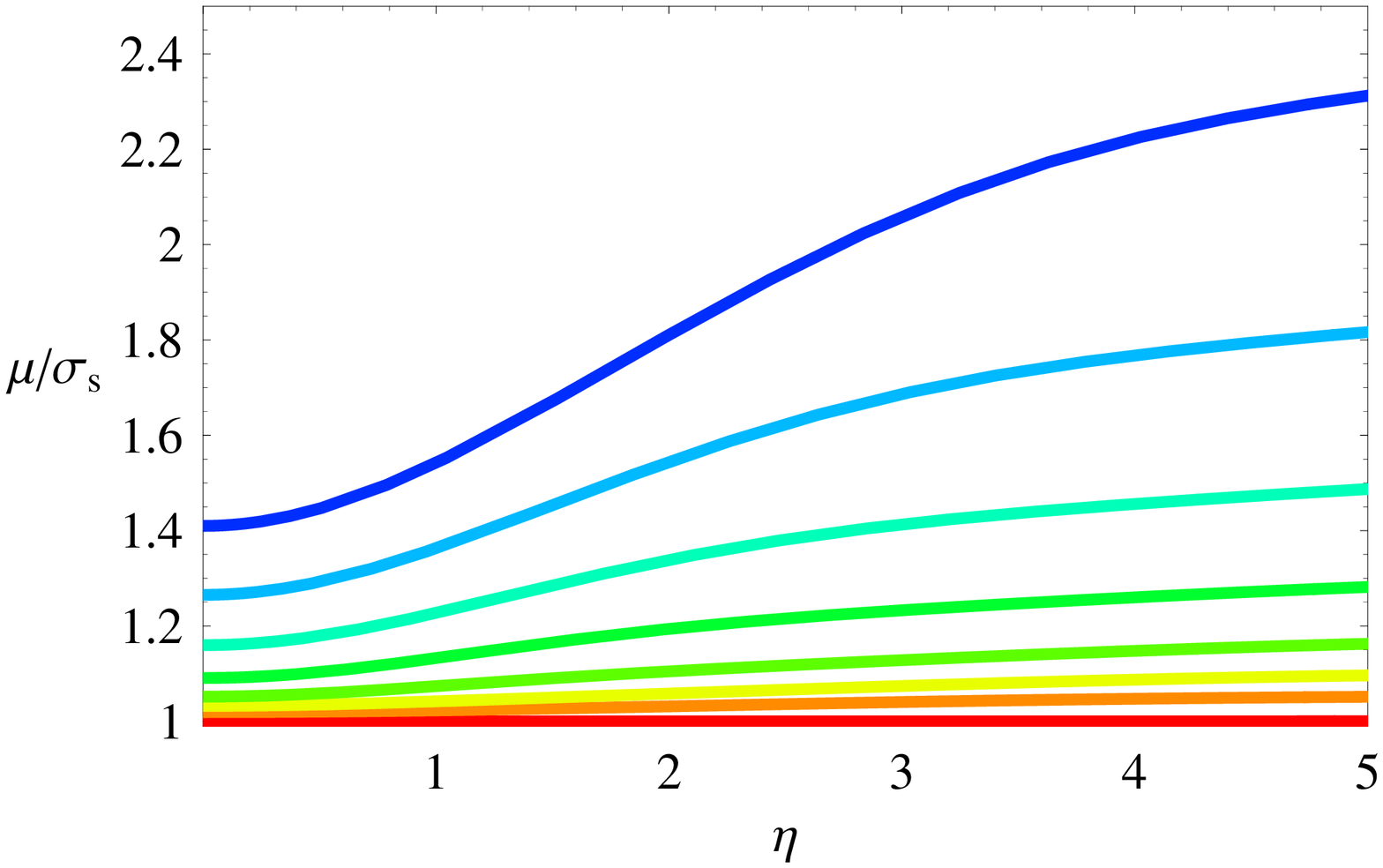}
 & \includegraphics[width=7.25cm]{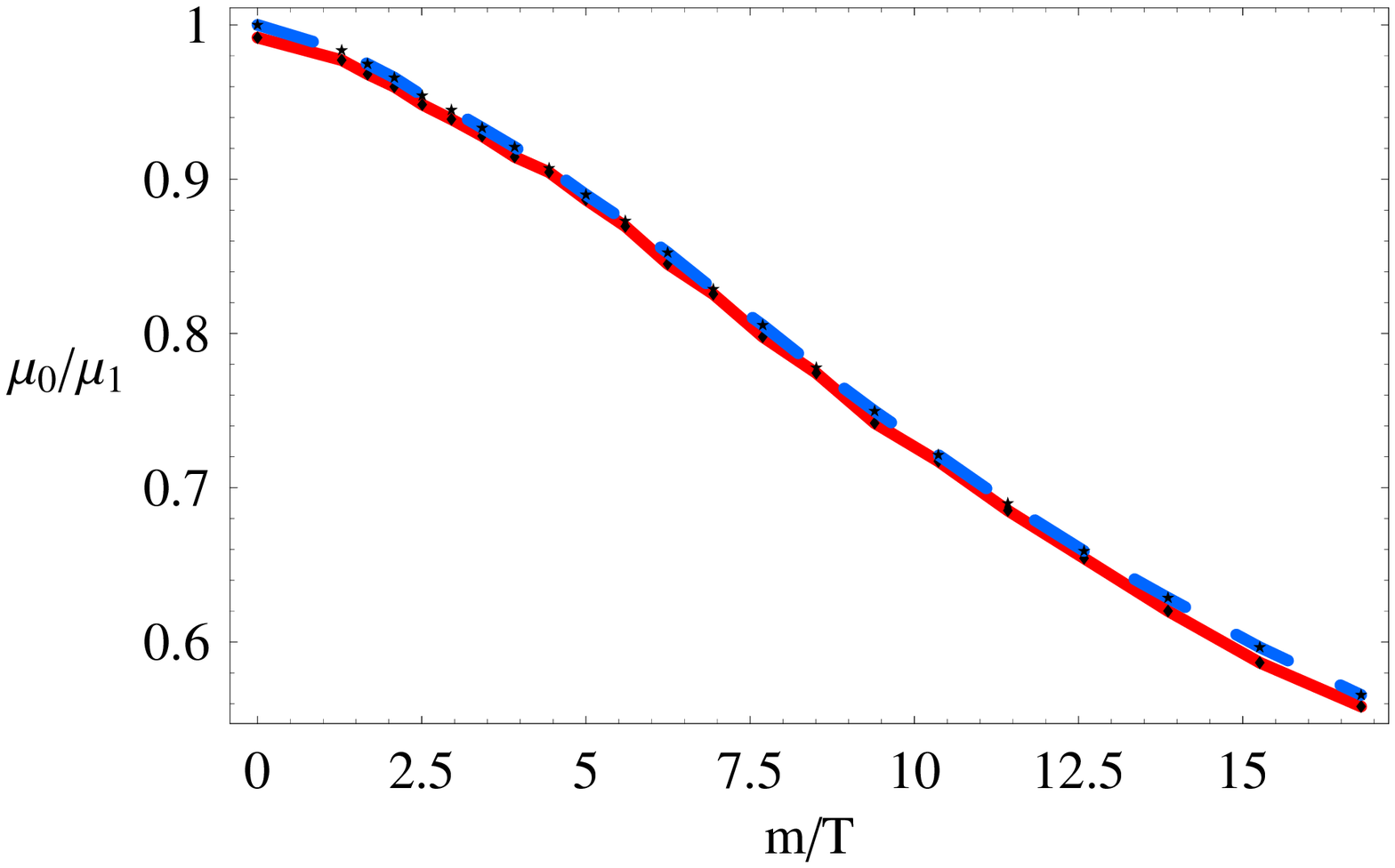} \\
 (A) & (B)
\end{tabular}
\caption{\label{dragfig}(A) Drag coefficient in string tension units as a function of the rapidity $\eta={\rm arctanh}\, v$, for $m/T=0, 2.08, 3.42, 5.00, 6.94, 9.40, 12.58, 16.80$ from bottom to top. (B) Ratio between the low and High velocity drag coefficients, as a function of $m/T$. The solid line corresponds to the direct evaluation using the string solution and the dashed line to the formula \eqref{relativedrag}.}
}

\subsection{Holographic computation of momentum broadening}\label{holomomtransf}

In the field theory side, the stochastic forces $\zeta_i^T$, $\zeta_i^L$ in the Langevin equation \eqref{langevin} correspond to operators built with heavy quark fields, so it is natural to identify them with small fluctuations of the string due to thermal radiation coming from the black hole \cite{Gubser:2006nz,CasalderreySolana:2007qw}. Notice that a force has to be applied to the quark to make it move at constant velocity, so the quark is not in thermal equilibrium with the plasma. However, the state described by the trailing string is a steady state with a constant momentum flow.

In the derivation of the momentum transfer, the worldsheet black hole is proposed to be a holographic description of the Wilson line spanned by the heavy quark, so the steady state is described effectively as a thermal state of a one-dimensional theory with temperature $T_{st}$. The worldsheet temperature depends on the velocity of the quark and it is in general different from the black hole temperature. At zero velocity the quark is at equilibrium with the plasma, so $T_{st}(v=0)=T$. Using a path integral approach, the degrees of freedom beyond the worldsheet horizon can be integrated out. The effect is to introduce a random force for the string endpoint at the worldsheet horizon that propagates to the boundary and introduces the stochastic forces of the Langevin description. This is equivalent to consider the Hawking radiation in the worldsheet theory. This introduces the picture of an ensemble of strings with fluctuating velocities. The large time average configuration is the classical string solution and the mean deviation is related to the magnitude of the random forces \cite{deBoer:2008gu,Son:2009vu,Giecold:2009cg,CasalderreySolana:2009rm}.

Using this approach, the value of the momentum transfer coefficient can be found from a Kubo formula involving the retarded Green's function of fluctuations of the string profile
\begin{equation}\label{momtransf}
\kappa_a=-\lim_{\omega\to 0} {2 T_{st}\over \omega} \,{\rm Im}\,G_R^{(a)}(\omega)\,, \ \ a=L,T\,.
\end{equation}

Consider small perturbations around the trailing string
\begin{equation}
x_1=vt+\xi(r)+\delta x_1(t,r)\ \ \ x_2=\delta x_2(t,r), \ \ \ x_3=\delta x_3(t,r)\,.
\end{equation}
Expanding to quadratic order, the Lagrangian for the fluctuations is\footnote{There are also first order terms but those are total derivatives, so they do not affect to the equations of motion and will be ignored here.}
\begin{equation}
\cL^{(2)}=-\cG_T^{\alpha\beta} \partial_\alpha \delta x_1 \partial_\beta \delta x_1-\sum_{i=1,2} \cG_L^{\alpha\beta} \partial_\alpha \delta x_i \partial_\beta \delta x_i\,,
\end{equation}
where
\begin{equation}
\cG_a^{\alpha\beta}={1\over 2} f_a \sqrt{-h} h^{\alpha\beta}\,,
\end{equation}
and
\begin{equation}
f_T=\Omega^2 e^{2A}\,,
\end{equation}
\begin{equation}\label{fas}
f_L={\Omega^4 e^{4A+2B}-\pi_\xi^2\over \Omega^2 e^{2 A}(e^{2 B}-v^2)}\,.
\end{equation}
The ratio $f_T/f_L$ for large values of $r$ becomes the right one for the zero temperature theory, $f_T = (1-v^2) f_L$. The velocity dependent factor corresponds to the anisotropy between the transverse and longitudinal fluctuations at short times or large frequencies. However, the small frequency behavior should be related to values of $r$ close to the worldsheet horizon $r_H$, determined by $e^{2 B}=v^2$. Expanding close to the horizon and using \eqref{pixi}, one finds
\begin{equation}\label{frH}
f_L(r_H)=f_T(r_H)\left(1+2{A'(r_H)\over B'(r_H)}+2{\Omega'(r_H)\over \Omega(r_H) B'(r_H)} \right)\,.
\end{equation}
In the $\cN=4$ theory this gives
\begin{equation}
f_L(r_H)={1\over (1-v^2)^{3/2}}, \ \ f_T(r_H)={1\over (1-v^2)^{1/2}}\,,
\end{equation}
so the zero temperature anisotropy persists at the worldsheet horizon.

The metric $h_{\alpha\beta}$ can be diagonalized\footnote{This is done by changing the time variable $t$ to $t=\tilde t+g(r)$, with $g'=-h_{tr}/h_{tt}$. After this change of variables, the new components $\tilde h_{\alpha\beta}$ of the metric are $\tilde h_{tt}=h_{tt}$, $\tilde h_{tr}=0$, $\tilde h_{rr}=h_{rr}-h_{tr}^2/h_{tt}$. In the text the new variables are relabeled to $\tilde t \to t$, $\tilde h\to h$.}
 $h_{\alpha\beta}=\,{\rm diag}\,( -\alpha, 1/\alpha)$ with
\begin{equation}
\alpha={ \left[(e^{2B}-v^2) (\Omega^4 e^{4A+2 B}-\pi_\xi^2) \right]^{1/2}\over \Omega^2 e^{A+B} }\,.
\end{equation}
Close to the horizon, the function $\alpha \simeq \alpha_0(r-r_H)$ vanishes
$$
\alpha\simeq \alpha_0(r-r_H)+\cdots =
$$
\begin{equation}
 \ \ \ =2 e^{A(r_H)+B(r_H)} \left[B'(r_H) \left(2{\Omega'(r_H)\over \Omega(r_H)}+2 A'(r_H)+B'(r_H) \right) \right]^{1/2} (r-r_H)+\cdots\,.
\end{equation}
Requiring that the Euclidean solution is regular at the horizon implies that the temperature of the worldsheet black hole is $T_{st}=\alpha_0/(4\pi)$.
The temperature at $v=0$ coincides with the black hole temperature $T_{st}=T$. In the $\cN=4$ theory, the worldsheet temperature has a simple dependence with the velocity $T_{st}=T/\sqrt{\gamma}$. The deviation from the conformal case has been represented in figure \ref{tst}.

\FIGURE[ht!]{
\includegraphics[width=10cm]{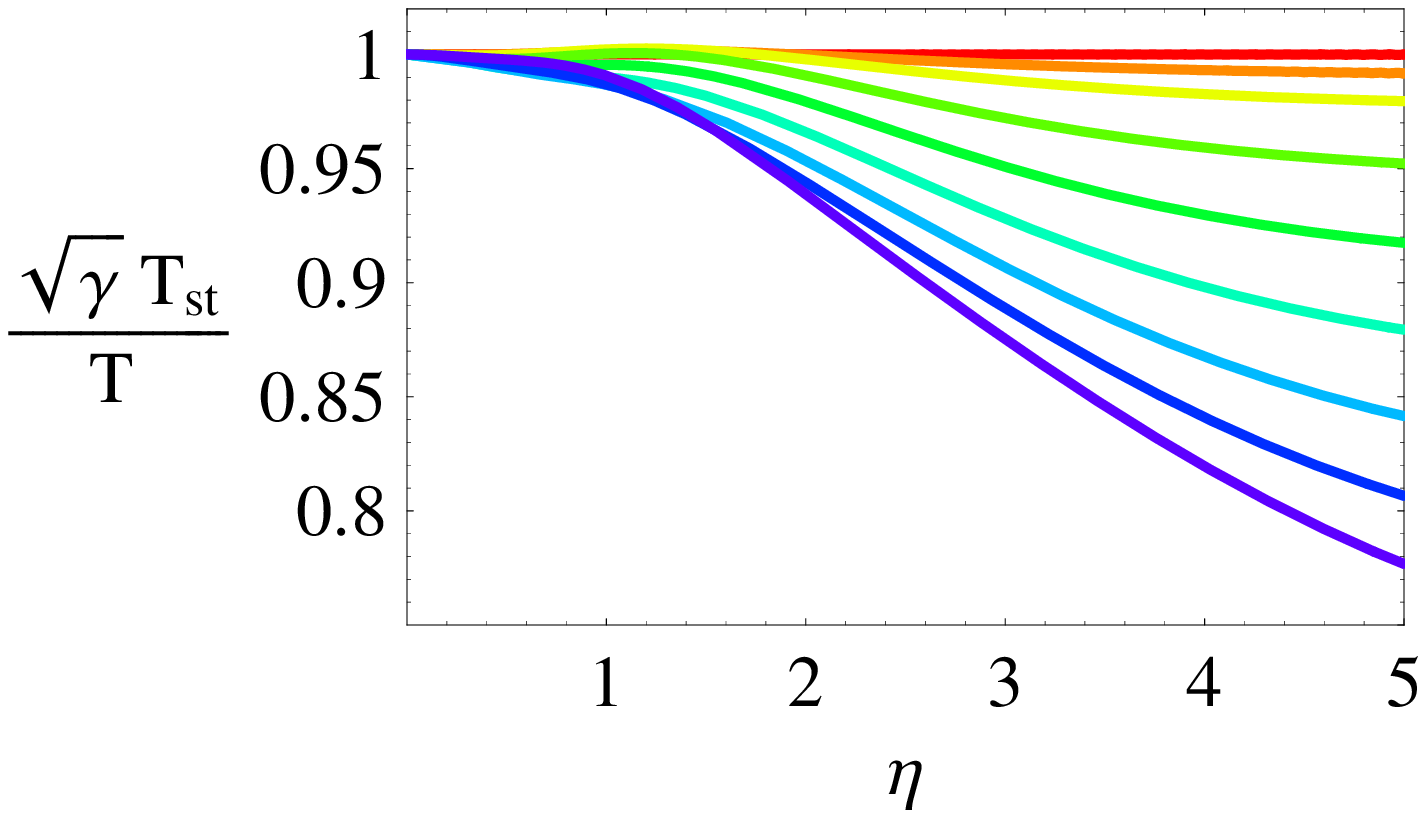}
\caption{\label{tst} Temperature of the worldsheet black hole as a function of the rapidity $\eta={\rm arctanh}\, v$, for $m/T=0, 1.28, 2.08, 3.42, 5.00, 6.94, 9.40, 12.58, 16.80$ from top to bottom.}
}

It is convenient to use a Fourier decomposition of the fluctuations. A Fourier mode $\delta x = e^{-i\omega t} \varphi(r)$ obeys the following equation ($a=T,L$)
\begin{equation}
\varphi_a''+(\log(\alpha f_a))' \varphi_a'+{\omega^2\over \alpha^2} \varphi_a=0\,.
\end{equation}
In order to compute a retarded correlator, the solution must be ingoing close to the horizon
\begin{equation}
\varphi_a = (r-r_H)^{-i\omega/\alpha_0} F_a(r)\, \ \ \ F_a(r_H)=\,{\rm const.}\,.
\end{equation}
One can use a low frequency expansion
\begin{equation}
F=F_a^{(0)}(r)+\omega F_a^{(1)}(r)+\cdots\,.
\end{equation}
The leading term is just a constant and can be fixed to $F_a^{(0)}(r)=1$. The equation for the next term can be simplified by choosing $W_a=-i {F_a^{(1)}}'$
\begin{equation}
W_a'+(\log(\alpha f_a))'W_a+{1\over \alpha_0(r-r_H)}\left({1\over r-r_H}- (\log(\alpha f_a))'\right) = 0\,.
\end{equation}
One can rewrite this as
\begin{equation}
\left[ \alpha f_a W_a\right]'-{1\over \alpha_0}\left[{\alpha f_a\over r-r_H} \right]'=0\,.
\end{equation}
Then, imposing regularity at the horizon $r\to r_H$, the solution is
\begin{equation}
W_a=-{f_a^H\over \alpha f_a}+{1\over \alpha_0(r-r_H)}\,,
\end{equation}
where $f_a^H=f_a(r_H)$.

In order to compute the Green's function a cutoff $r_\Lambda\gg L, r_H$ is introduced and the normalization of the solutions is fixed at the cutoff $Y_a(\omega,r_\Lambda)=e^{-i\omega t}$, so
\begin{equation}
Y_a(\omega,r)=e^{-i\omega t}{\varphi_a(r)\over \varphi_a(r_\Lambda)}\,.
\end{equation}
The retarded Green's function is proportional to the boundary action as the cutoff is taken to infinity. To leading order in $\omega$, the result is
\begin{equation}\label{green}
G_R^{(a)}(\omega)={1\over 2\pi\alpha'}\lim_{r_\Lambda\to \infty} f_a \sqrt{-h} h^{rr} Y_a^*(\omega,r_\Lambda)\partial_r Y_a(\omega,r_\Lambda) = - i\omega {f_a^H\over 2\pi\alpha'}+\cO(\omega^2)\,.
\end{equation}
Using \eqref{frH} and \eqref{green} in \eqref{momtransf}, the value of the transverse momentum transfer is
\begin{equation}\label{kappa}
\kappa_T= 2 \gamma T_{st} \mu\,.
\end{equation}
The longitudinal momentum transfer can be simplified using that $v=e^{B(r_H)}$ and
\begin{equation}
{\partial\tilde\mu\over \partial v} \equiv \left.{\left(\partial \tilde\mu/ \partial r\right)\over  \left(\partial e^B/ \partial r\right)}\right|_{r=r_H}  ={\tilde\mu(r_H)\over v}\left(-\gamma^2 v^2+2{\Omega'(r_H)\over\Omega(r_H) B'(r_H)}+2{A'(r_H)\over B'(r_H)} \right)\,.
\end{equation}
The complicated dependence on the derivatives of the functions of the metric is the same as in \eqref{frH}. This allows to write the value of the longitudinal momentum transfer as
\begin{equation}
\kappa_L =2\gamma^3 T_{st}\mu + 2\gamma T_{st} v{\partial \mu \over \partial v} \,.
\end{equation}
There is an additional term due to the velocity dependence of the drag in the non-conformal theory, and the velocity dependent factor in the conformal theory is $\gamma^3$ instead of the factor $\gamma$ found for the transverse momentum transfer. In the following an interpretation for this difference is given.

Assume that it is possible to describe an ensemble of heavy quarks using a Liouville equation. The probability density at time $t$ of finding a heavy quark in the position $x$ and with transverse and longitudinal momenta $p_T$ and $p_L$ is $\rho(x,p_T,p_L;t)$ \footnote{All these quantities are vectors but arrows are dropped for notational simplicity.}. The continuity equation is
\begin{equation}\label{liouville}
{\partial \rho \over \partial t}+{\partial \over \partial x} (\dot{x} \rho) + {\partial \over \partial p_T} (\dot{p_T} \rho)+{\partial \over \partial p_L} (\dot{p_L} \rho)  =0\,.
\end{equation}
The trailing string describes a quark moving through a viscous medium and subject to an external force in the longitudinal direction $F_v= \gamma v \mu$
\begin{equation}
\dot{p_L} = F_v-\mu {p_L\over M}\,, \ \ \dot{p_T}=-\mu {p_T\over M}\,.
\end{equation}
The effect of the stochastic forces can be taken into account adding to the right-hand side of \eqref{liouville} terms of the form ${\kappa\over 2} {\partial^2\rho/ \partial p^2}$. In the absence of forces $F_v=0$ and $\mu=0$, there is a homogeneous solution $\rho \sim (\kappa t)^{-3/2} e^{-p^2/(2 \kappa t)}$ to \eqref{liouville} that describes correctly the spreading of the quarks due to Brownian motion.

The question is now if there is a static and homogeneous solution to \eqref{liouville}. This will mean that the heavy quarks and the plasma are in a steady state configuration with a constant flux of momentum, introduced in the system by the external force acting on the quarks. When there is no force the average velocity of the quarks vanish and the system is at thermal equilibrium. Under these conditions, the equations reduce to
$$
\mu {p_T\over M} \rho+{\kappa_T\over 2} {\partial\rho \over \partial p_T} =0\,,
$$
\begin{equation}
-\gamma v\mu+\mu {p_L\over M} \rho+{\kappa_L\over 2} {\partial\rho \over \partial p_L} =0\,.
\end{equation}
There is indeed a solution in the form of a Boltzmann distribution centered at non-zero longitudinal momentum
\begin{equation}\label{rhoeq}
\rho_{\rm eq}(p_T,p_L) \propto \exp\left(-{p_T^2\over 2M \gamma T_{st}}\right)  \exp\left(-{(p_L-\gamma v M)^2\over 2M \gamma^3 T_{st}+2 M T_{st} v\partial_v \mu}\right)\,.
\end{equation}
Focusing on the conformal case, the factor $\gamma^2$ can be understood as a consequence of the underlying Lorentz symmetry. By doing a Fourier transformation, the resulting distribution in the spatial coordinates has widths
\begin{equation}\label{widths}
\vev{\Delta x_T^2}\sim {1\over \gamma T_{st} M} \equiv L^2, \ \ \vev{\Delta x_L^2}\sim {1\over \gamma^3 T_{st} M} = {1\over \gamma^2} L^2\,.
\end{equation}
The square roots of the widths $\langle{\Delta x_{T,L}^2}\rangle^{1/2}$ give the mean length of the distribution along the transverse and the longitudinal directions in real space, so under a longitudinal boost the transverse width should not change and the (square root) of the longitudinal width should suffer the usual Lorentz contraction. If the distribution of the ensemble of heavy quarks is spherical with mean radius $L$ in the rest frame defined by the average values $\vev{x_T}=0$, $\vev{x_L}=v t$, then the boost to the rest frame of the plasma will account for the different $\gamma$ factors in the longitudinal and transverse widths in \eqref{widths}. The possible relation of the additional $\gamma^2$ factor in the longitudinal direction with Lorentz invariance was pointed out in \cite{Gubser:2006nz} from the analysis of the zero temperature worldsheet Green's function. Notice that there is still a dependence of the widths on the modulus of the relative velocity between the heavy quarks and the plasma, implicit in $T_{st}$.

In the non-conformal case, in addition to the Lorentz contraction, the velocity dependence of the drag coefficient affects to the distribution in the longitudinal direction, so it is not longer spherical in the rest frame of the quark distribution. This is even more explicit if the drag coefficient is taken as a function of the average longitudinal momentum $P_L=\gamma v M$. Then, the longitudinal momentum transfer has a rather simple form
\begin{equation}
\kappa_L = 2\gamma^3 T_{st} \mu\left(1+ {P_L\over \mu} {\partial \mu\over \partial P_L}\right)\,.
\end{equation}
A measure of the anisotropy in momentum space as a function of the velocity for different masses in the $\cN=2^*$ theory can be found in figure \ref{anisotfig}.

\FIGURE[ht!]{
\includegraphics[width=11cm]{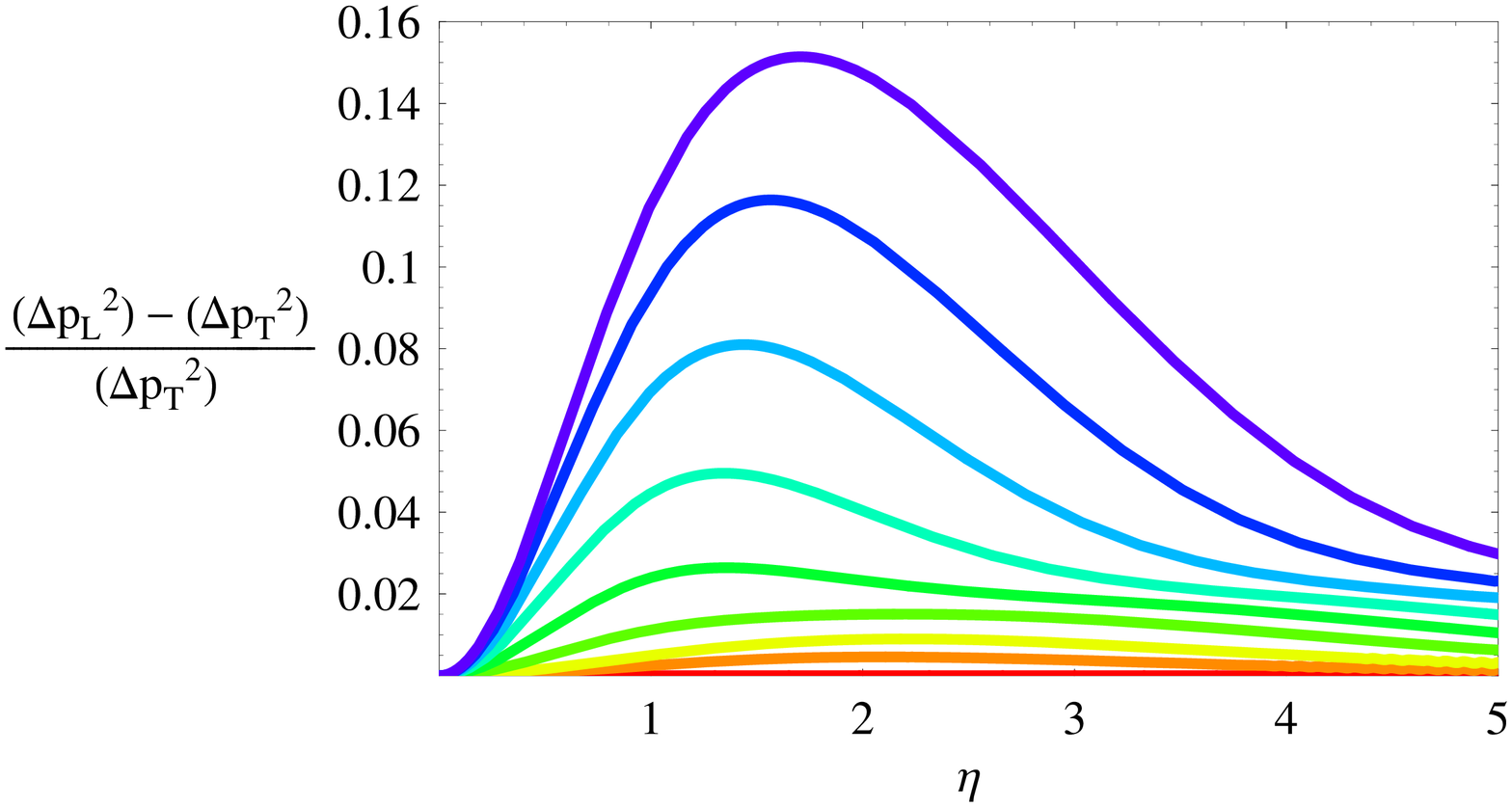}
\caption{\label{anisotfig} Anisotropy of the momentum distribution of heavy quarks in the rest frame of the ensemble as a function of the rapidity $\eta={\rm arctanh}\, v$, for $m/T=0, 1.28, 2.08, 3.42, 5.00, 6.94, 9.40, 12.58, 16.80$ from bottom to top.}
}

\section{Jet quenching}\label{jetsec}

A different estimation of the energy loss of relativistic hadrons in the plasma \cite{Wiedemann:2000za} is based on a weak coupling calculation of the gluon radiation emitted by a dipole source in the limit $E\gg p_T\gg T$, where $p_T$ is the typical transverse momentum of the radiated gluons and $E$ is the energy of the source. The momentum transfer to the medium $\hat q$, known as `jet quenching' parameter, can be associated to a rectangular adjoint Wilson loop with two lightlike sides of length $L_-$, related to partons moving at relativistic velocities, and two spatial sides of length $L_x\ll L_-$ related to the transverse momentum of the radiated gluons $p_T\sim 1/L_x$. In the planar limit, the value of the adjoint Wilson loop can be obtained from a Wilson loop in the fundamental representation
\begin{equation}
\vev{W_{\rm Adj}(\cC)} = \left| \vev{W(\cC)}\right|^2\,,
\end{equation}
In \cite{Liu:2006ug} it was proposed to use the expectation value of the fundamental Wilson loop as a definition for the momentum transfer $\hat q$ at strong coupling
\begin{equation}\label{jetloop}
\vev{W(\cC)} = e^{-{1\over 8\sqrt{2}} \hat q L_x^2L_-}\,,
\end{equation}
and, using a classical string in the $AdS_5$ black hole, the following result for the $\cN=4$ theory was found
\begin{equation}\label{quenchN4}
\hat q_{\cN=4} = {\pi^{3/2} \Gamma\left({3\over 4}\right) \over \Gamma\left( {5\over 4}\right)} \sqrt{\lambda} T^3\,.
\end{equation}
One can see that once the dependence of the coupling and the temperature have been factored out, the jet quenching differs from the momentum transfer obtained from worldsheet fluctuations of the trailing string \eqref{kappa}. The relation between the jet quenching and the drag will be made explicit below.

The expectation value of the Wilson loop \eqref{jetloop} can be computed evaluating the classical action $S(\cC)$ of a string ending on the contour $\cC$ at the boundary.
The contour $\cC$ is chosen to be rectangular in the $y^-=(t-y)/\sqrt{2}$ and $x$ directions, with the length in the $y^-$ direction $L_-$ very large, so it is possible to neglect any dependence on the $y^-$ coordinate. The choice of coordinates for the associated string is $\tau=y^-$, $r=\sigma$ and the profile is determined by the function $x(r)$. The Lagrangian is
\begin{equation}
\cL= {\Omega^2 e^A\over \sqrt{2}} \sqrt{1-e^{2 B}}\sqrt{1+e^{2 A} {x'}^2}\,.
\end{equation}
From this one can derive the equation of motion
\begin{equation}
x' = {c e^{-A}\over \sqrt{\Omega^4 e^{4 A}(1-e^{2 B})-c^2}}\,.
\end{equation}
This can be simplified by defining a `drag coefficient' function as
\begin{equation}
\tilde \mu(r)=(1-e^{2 B(r)})^{1/2}\Omega(r)^2 e^{2 A(r)}\,.
\end{equation}
Notice that $\tilde \mu(r)$ at fixed $r$ coincides with the drag coefficient \eqref{dragcoef} (up to the $2\pi\alpha'$ factor) at velocity $v^2= e^{2 B(r)}$. The radial coordinate is related to the energy scale in the dual theory, so there is a scale associated to each velocity. The map to a physical scale in the field theory could be ambiguous, depending on the choice of radial coordinate. A better option could be to use the energy scale defined by the average transverse kinetic energy of the heavy quark, using the momentum transfer \eqref{kappa} and assuming a non-relativistic dispersion relation as in \eqref{rhoeq}
\begin{equation}
\vev{E_T}\equiv {\vev{p_T^2}\over 2 M} ={\kappa_T\over 2 \mu}= \gamma T_{st}\,.
\end{equation}

The string profile ends at a finite value of the radial coordinate $r>0$ unless the integration constant $c^2\leq \tilde \mu^2(0)=\tilde\mu_0^2$. This implies that in the non-conformal case there are two branches. In the $c^2>\tilde \mu_0^2$ branch the action of the string grows linearly with the length $L_x$ in the $x$ direction, while in the $c^2<\tilde \mu_0^2$ branch it grows quadratically. A similar behavior has been observed previously \cite{Argyres:2008eg}. A numerical calculation in the $\cN=2^*$ case shows that the branch with $c^2> \tilde\mu_0^2$ has a larger action for a given length, so it is disfavored. In the following only $c^2\leq \tilde\mu_0^2$ will be considered.

Notice that $x'\neq \infty$ at the horizon. In principle this looks like the solution is irregular or should be continued beyond the horizon. However, it is possible to see that the solution corresponds to a regular string touching the horizon once Schwarzchild coordinates are used. The Gaussian radial coordinate $r$ is related to the Schwarzschild radial coordinate $u$ as $dr = d u/\sqrt{g(u)}$, where $g(u_H)=0$ at the horizon $u_H>0$. Then,
\begin{equation}
x'={d x\over d r} = \sqrt{g(u)} {d x\over d u}\,.
\end{equation}
If $x'$ is constant at the horizon the derivative with respect to the Schwarzschild coordinate diverges ${d x\over d u}\to \infty$, and this gives the right condition for a regular solution touching the horizon.

With this choice, the length in the $x$ direction and the regularized action of the string are
$$
{L_x\over 2} =\int_0^\infty dr\; {e^{-A}c \over \sqrt{\tilde\mu^2-c^2}}\,,
$$
\begin{equation}
{S\over 2 } = {L_-\over 2\sqrt{2} \pi\alpha'}\int_0^\infty dr\; e^{-A}\tilde \mu\left[{\tilde \mu\over \sqrt{\tilde\mu^2-c^2}}-1\right]\,.
\end{equation}
The jet quenching is obtained from the $L_x\to 0$ behavior, this corresponds to the $c\to 0$ limit in the integrals.
$$
{L_x\over 2}\simeq c \int_0^\infty dr {e^{-A}\over \tilde\mu}\,,
$$
\begin{equation}
 {S\over 2}\simeq {c^2 L_-\over 4\sqrt{2}\pi\alpha'} \int_0^\infty dr {e^{-A}\over \tilde\mu}\,.
\end{equation}
So the (inverse of) the jet quenching coefficient is
\begin{equation}\label{jetdrag}
\hat q^{-1} =\pi\alpha'  \int_0^\infty dr {e^{-A}\over \tilde \mu}\,.
\end{equation}
Taking into account the relation between the radial coordinate and the energy scale in the dual theory, the formula above could be interpreted as an average of the drag over all scales, showing a relation between the drag and the jet quenching. Using the relation between the scale and the velocity $v=e^{B(r)}$, \eqref{jetdrag} could also be interpreted as an average over velocities, although the explicit form is more complicated.  The value of $\hat q$ in units of the spatial string tension and temperature are plotted in figure \ref{jetfig} for different masses in the $\cN=2^*$ theory.
One can see that the jet quenching grows with $m/T$, the change is about $\sim$ 50\% for $m/T\sim 17$. The numerical value at $m/T=0$ $\hat q\simeq 4.794\, T\,\sigma_s$ is in good agreement with the analytical result \eqref{quenchN4}
\begin{equation}
\hat q_{\cN=4} = {2 \sqrt{\pi} \Gamma\left({3\over 4}\right) \over \Gamma\left( {5\over 4}\right)} T\,\sigma_s \simeq 4.793 \, T\,\sigma_s\,.
\end{equation}

\FIGURE[ht!]{
\includegraphics[width=10cm]{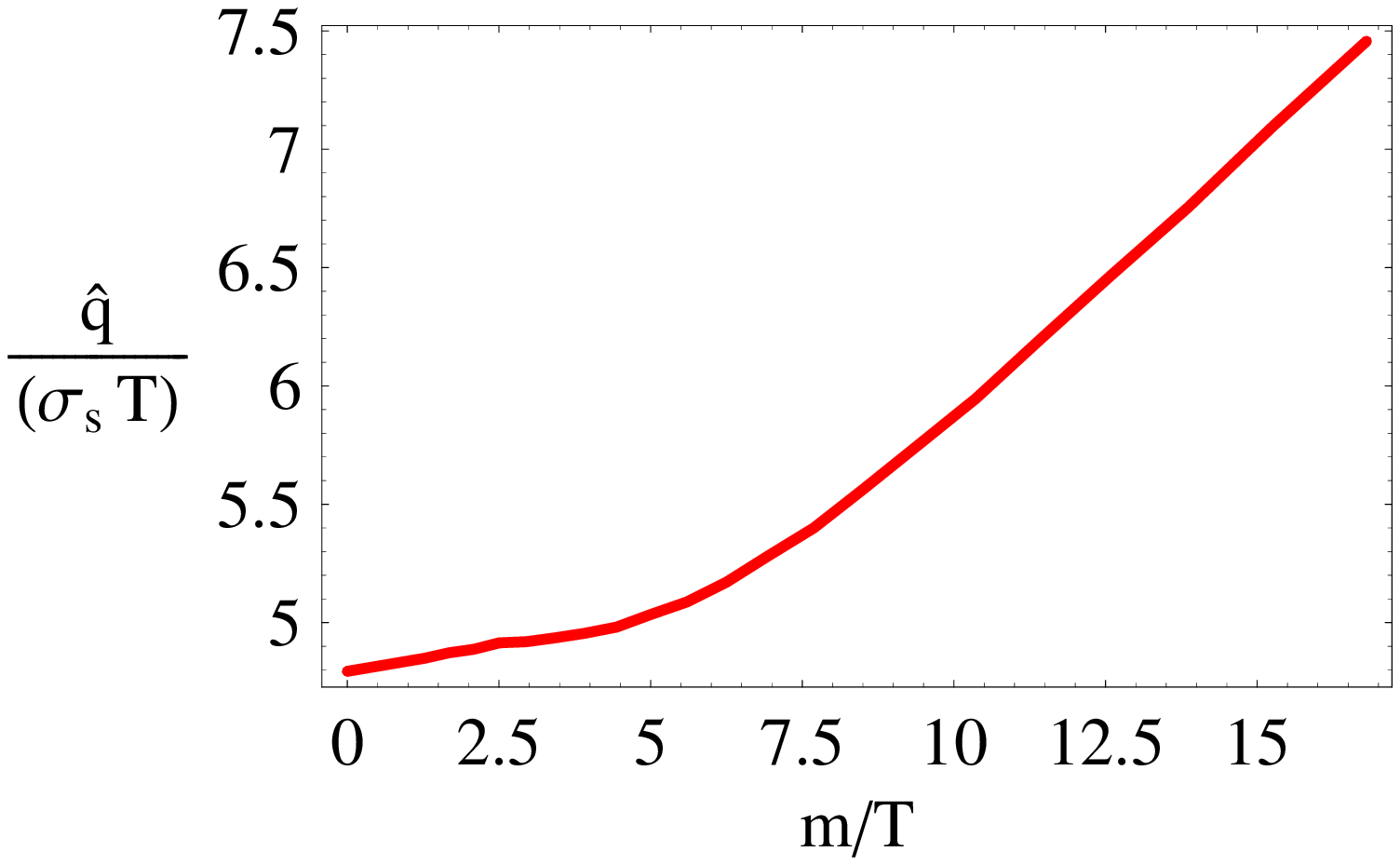}
\caption{\label{jetfig}Jet quenching in $\sigma_s T$ units as a function of $m/T$.
}}

\section{Discussion of the results}\label{concsec}

In \cite{Horowitz:2007su} some experimental observables for heavy quarks were proposed such that it would be possible to make comparisons with perturbative QCD predictions. A matter of concern is how the comparison with `real world' physics should be made. There are two parameters that enter explicitly in both the drag and the jet quenching, the 't Hooft coupling and the temperature\footnote{In the $\cN=2^*$ theory the mass deformation is an additional parameter that does not enter explicitly but that modifies the value.}. There have been different proposals to fix their value, for instance by fixing some physical quantity in both theories, like the energy density or the diffusion coefficient e.g. \cite{Gubser:2006qh,CasalderreySolana:2006rq}.
However, different choices of the observables used to do the matching will lead to different quantitative results. Besides, it is difficult to quantify the value of $m/T$ that would lead to the best comparison with QCD. A rough estimation of the deviation from conformal invariance that corresponds to QCD would be to compare the speed of sound or the equation of state of both theories\footnote{I would like to thank Krishna Rajagopal and Hong Liu for suggesting these comparisons.}. The speed of sound is the most straightforward, although it would be desirable to use the equation of state as well to check its robustness. Using the values obtained in the lattice \cite{latticeEOS} and in the $\cN=2^*$ theory \cite{Buchel:2007vy}, one finds that for $T\sim 1.5-2\; T_c$ the value of $m/T$ is in the approximate range $3\lesssim m/T \lesssim 6$. The uncertainty of this comparison motivates the study of quantities within the same theory, as has been done for the transport coefficients of the plasma, the shear to entropy density ratio being the most obvious example.

The expression \eqref{mu1} shows that the ultrarelativistic drag coefficient is sensitive to the infrared physics, and that it should be related to transport coefficients of the plasma obtained from the energy momentum tensor. One can also see from the formulas \eqref{stringt} and \eqref{relativedrag} that a smaller spatial string tension in the non-conformal theory implies a relatively larger drag at high velocities and vice versa. A relation between the jet quenching and the drag coefficient is given in section \ref{jetsec}, in the formula \eqref{jetdrag} one can see that the value of the jet quenching is an average of the drag coefficient over all scales. Therefore, their values are correlated as should be expected.

The numerical results for the $\cN=2^*$ theory presented in figures \ref{dragfig} and \ref{jetfig} show that the energy loss is larger in string tension units than in the non-conformal case, and grows as the mass deformation becomes more important. For the range of masses studied here $0\leq m/T \lesssim 17$, the jet quenching changes up to $\sim$ 50\%. The drag coefficient can be more than twice as large as the conformal value at high velocities.

In addition to the drag coefficient, the momentum broadening of heavy quarks \cite{Gubser:2006nz,CasalderreySolana:2007qw} was studied in section \ref{holomomtransf}. The interpretation of the original results is much less clear in this case, since the values that were obtained did not fit with the expectations from a Langevin description of the dynamics of heavy quarks at thermal equilibrium with the plasma. A recent work \cite{Giecold:2009cg}, gives a possible interpretation in terms of energy loss through non-thermal radiation processes.
The transverse and longitudinal momentum transfer coefficients are found to be
\begin{equation}
\kappa_T= 2 \gamma T_{st} \mu, \ \ \kappa_L=2\gamma^3 T_{st} \mu \left(1+{P_L\over \mu} {\partial \mu \over\partial P_L}\right)\,.
\end{equation}
where $P_L$ is the average longitudinal momentum of the heavy quark. It is argued that these formulas are compatible with steady state configurations of quarks subject to an external force and moving through a relativistic viscous medium.
The additional $\gamma^2$ factor in $\kappa_L$ relative to $\kappa_T$ is argued to be due to Lorentz contraction in the longitudinal direction, under the assumption that the dynamics of heavy quarks can be described with a Liouville equation, at least as a first approximation.

In the conformal case this implies that the distribution in momentum will be spherical in the rest frame of the quark ensemble, but in the non-conformal case there is an anisotropy. It is surprising that the anisotropy of the system does not introduce an anisotropy in the distribution in the conformal case. It would be interesting to understand whether this can be derived from conformal invariance or if it is a characteristic of the holographic computation. It is also possible that, even in a first approximation, the dynamics of the heavy quarks are described by something more complicated than a simple Liouville equation, in such a way that the simple relation of the $\gamma^2$ factor to a Lorentz transformation does not hold.

In the $\cN=2^*$ theory a broadening of the longitudinal momentum is observed (Fig.~\ref{anisotfig}), related to the growth of the drag coefficient at larger velocities. A broadening of the momentum distribution in the longitudinal direction has also been observed in perturbative calculations of quarks moving through an anisotropic plasma \cite{Romatschke:2006bb}. This seems to be the right qualitative behavior that is consistent with the observed distribution of jets in the plasma. It would be interesting to do a comparison with the broadening predicted by holographic models.

\section*{Acknowledgments}
I would like to thank P.~Romatschke and D.~Teaney for illuminating discussions on the drag force and momentum broadening of the heavy quarks. I would want to thank also B.~Bringoltz and A.~Karch for useful comments. Special thanks go to S.~Paik and L.~Yaffe for their valuable comments on the draft. I want also to acknowledge very interesting discussions with E.~Kiritsis and H.~Liu and K.~Rajagopal.
This work was supported in part by the U.S. Department of Energy under Grant No. DE-FG02-96ER40956.

\begin{appendix}
\section{Numerical method}\label{app}

The numerical calculations are made following the conventions of \cite{Buchel:2007vy,Buchel:2003ah}. The Einstein equation \eqref{einsteineqs} together with the equations of motion for the scalar fields \eqref{scalareqs} form a system of second order ODEs with four independent equations. The radius of curvature is fixed to $L =2$ to make the gauged supergravity coupling unity. The leading terms in the asymptotic expansion of the scalar fields as $r\to \infty$ are
\begin{equation}
\chi \simeq \chi_{\infty} e^{-r/2}+\cdots, \ \ {\alpha\over \sqrt{3}} \simeq -\rho_{\infty} r e^{-r}+\cdots
\end{equation}
From \eqref{scalarboundexp}, solutions with $\cN=2$ supersymmetry should satisfy the condition
\begin{equation}\label{susycond}
\left| {\chi_{\infty}^2\over 3 \rho_{\infty}}-1\right| = 0
\end{equation}
The asymptotic behavior of the metric functions should be the same for all solutions. From the holographic dual perspective this means that they flow to the same ultraviolet fixed point. Comparing to the usual $AdS_5$ black hole, as $r\to \infty$,
\begin{equation}\label{boundcond}
e^{A(r)}\to e^{r/2}, \ \ e^{B(r)}\to 1.
\end{equation}
This fixes completely the boundary conditions at infinity for a fixed value of the mass.

Regularity of the solution at the horizon imposes additional constraints as $r\to 0$, namely
\begin{equation}
e^{B(r)} \simeq b_0 r +\cdots, \ \ A'(0)=0, \ \ \alpha'(0)=0, \ \  \chi'(0)=0\,.
\end{equation}
It will be convenient to extract the non-analytic behavior of the function $B(r)$ as
\begin{equation}
B(r)= 4 b(r)-4 A(r)+\log(b_0 r), \ \ b'(0)=0\,.
\end{equation}
It is actually easier to implement these conditions at the horizon and shoot toward the boundary, this is the approach used here. The conditions \eqref{boundcond} can be imposed by shifting the functions $A(r)$ and $B(r)$ afterwards. The condition \eqref{susycond} is used as a criterion to keep the solution as the space of initial values $\chi(0)=\chi_0$ and $\alpha(0)=\alpha_0$ is explored \footnote{I would like to thank Steve Paik for providing a set of initial values that helped very much for this analysis.}. In the numerical solutions, the value of the supersymmetric condition is always kept under $10^{-2}$. With this condition, the numerical solutions have a small relative mass difference between fermions and bosons
\begin{equation}
\left|{m_b-m_f\over m_b} \right|\lesssim 0.005\,.
\end{equation}

The shooting is done in Mathematica using a set of six first order differential equations and a second order differential equation. The choice is inspired by the equations of the supersymmetric flow \eqref{susyeqs}. Following those, the deviation from extremality is parameterized by three functions $a(r)$, $c(r)$ and $b_1(r)$ that vanish as the non-extremal flow approaches the supersymmetric one when $r\to \infty$,
\begin{equation}
\alpha' ={1\over 4}{\partial W \over \partial \alpha} + a, \ \
\chi'={1\over 4}{\partial W \over \partial \alpha} + c, \ \
b'= {W\over 3}+b_1,
\end{equation}
where $W$ is the superpotential \eqref{superpotential}. The Einstein equations and the equations of motion for the scalar fields give
\begin{equation}
r a' +a\left(1+4 r b_1-{4\over 3} r W +{1\over 4} r {\partial^2 W \over \partial \alpha^2}\right) +{1\over 4}{\partial W \over \partial \alpha}+r b_1 {\partial W \over \partial \alpha}+ r c {1\over 4}{\partial^2 W \over \partial \alpha \partial \chi}=0\,,
\end{equation}
\begin{equation}
r c' +c\left(1+4 r b_1-{4\over 3} r W +{1\over 4} r {\partial^2 W \over \partial \chi^2}\right) +{1\over 4}{\partial W \over \partial \chi}+r b_1 {\partial W \over \partial \chi}+ r a {1\over 4}{\partial^2 W \over \partial \alpha \partial \chi}=0\,,
\end{equation}
\begin{equation}
r b_1'+b_1\left(2+4 r b_1-{8\over 3} r W \right)-{2\over 3} W -{1\over 3} r c {\partial W \over \partial \chi}-{1\over 3} r a {\partial W \over \partial \alpha}=0\,.
\end{equation}
Finally, there is a second order equation that can be solved once the solutions to the first order system are known
\begin{equation}
rA''+A'-2 b'+4r A' b'+4 r{b'}^2-rb_1'+{1\over 3} r\left( {\partial W \over \partial \chi}\chi'+{\partial W \over \partial \alpha}\alpha'\right)=0\,.
\end{equation}
The boundary values at $r=0$ of the extra functions are fixed by the requirement of regularity
\begin{equation}
a(0)=-{1\over 4}\left.{\partial W \over \partial \alpha}\right|_{\alpha=\alpha_0,\chi=\chi_0}, \ \ c(0)=-{1\over 4}\left.{\partial W \over \partial \chi}\right|_{\alpha=\alpha_0,\chi=\chi_0}, \ \ b_1(0)=\left.{W\over 3}\right|_{\alpha=\alpha_0,\chi=\chi_0}\,.
\end{equation}
There are only two more boundary conditions that need to be fixed
\begin{equation}
b(0)=0, \ \ A(0)=0\,.
\end{equation}
This choice implies that the solutions will correspond to different temperatures.

\end{appendix}

\end{document}